\documentclass[prl,twocolumn,showpacs, showkeywords,amsmath,amssymb,superscriptaddress,nofootinbib]{revtex4-1}

\usepackage{amssymb}
\usepackage{amsmath}
\usepackage{epsfig}
\usepackage{epstopdf}
\usepackage{hyperref}
\usepackage{breakurl}
\usepackage{xcolor}
\usepackage{dsfont,bbm}

\usepackage{extarrows}
\usepackage{shuffle}

\usepackage{graphicx}
\usepackage{subfigure}

\begin{document}

\title{Is Yang-Mills Theory Unitary in Fractional Spacetime Dimensions?}

\author{Qingjun Jin}
\email{qjin@gscaep.ac.cn}
\affiliation{Graduate School of China Academy of Engineering Physics, No.~10 Xibeiwang East Road, Haidian District, Beijing, 100193, China}
\author{Ke Ren}
\email{renk9@mail.sysu.edu.cn}
\affiliation{School of Physics and Astronomy, Sun Yat-Sen University, Zhuhai 519082, China}
\affiliation{CAS Key Laboratory of Theoretical Physics, Institute of Theoretical Physics, Chinese Academy of Sciences,  Beijing, 100190, China}
\author{Gang Yang}
\email{yangg@itp.ac.cn}
\affiliation{CAS Key Laboratory of Theoretical Physics, Institute of Theoretical Physics, Chinese Academy of Sciences,  Beijing, 100190, China}
\affiliation{School of Fundamental Physics and Mathematical Sciences, Hangzhou Institute for Advanced Study, UCAS, Hangzhou 310024, China}
\affiliation{International Centre for Theoretical Physics Asia-Pacific, Beijing/Hangzhou 310024, China}
\author{Rui Yu\vspace{2mm}}
\email{yurui@imu.edu.cn}
\affiliation{School of Physical Science and Technology, Inner Mongolia University, Hohhot 010021, China}
\affiliation{Beijing Computational Science Research Center, Beijing 100193, China}
\affiliation{CAS Key Laboratory of Theoretical Physics, Institute of Theoretical Physics, Chinese Academy of Sciences,  Beijing, 100190, China}

\begin{abstract}

We present concrete evidence that Yang-Mills theory exhibits non-unitarity in non-integer spacetime dimensions. This violation of unitarity stems from evanescent operators that, while vanishing in four dimensions, are non-zero in general $d$ dimensions. We demonstrate that these evanescent operators lead to the emergence of both negative-norm states and complex anomalous dimensions. 
\end{abstract}


\maketitle

\section{Introduction}

\noindent
While the real world exists in four spacetime dimensions, we often need to consider quantum field theories (QFTs) in non-integer dimensions.
One example is the dimensional regularization method, in which the spacetime is analytically continued to $d=4-2\epsilon$ dimensions~\cite{tHooft:1972tcz}.
Another example is the $\epsilon$-expansion method used to compute critical exponents \cite{Wilson:1971dc},
where theories at Wilson-Fisher (WF) fixed points are defined in fractional dimensions depending on $\epsilon$.
Additional studies of QFTs in non-integer dimensions can be found in \emph{e.g.}~\cite{Parisi:1978mn, Svozil:1985ha, Eyink:1989dv, Kroger:2000wa, Calcagni:2011sz, Braun:2018mxm, Giombi:2019upv}.

Unitarity, as a fundamental physical assumption, plays a key role in many QFT studies.
Surprisingly,
it has been shown that the $\phi^4$ scalar field theory in non-integer dimensions is not unitary \cite{Hogervorst:2014rta, Hogervorst:2015akt}.
The source of unitarity violation is the so-called evanescent operators, which are non-zero in general $d$ dimensions but vanish as $d\rightarrow 4$.
It has been shown that
these evanescent operators give rise to negative-norm states
and also lead to complex anomalous dimensions \cite{Hogervorst:2014rta, Hogervorst:2015akt}.
Negative-norm states were also found in the Gross-Neveu-Yukawa theory \cite{Ji:2018yaf}.
An important question is: does a similar mechanism for unitarity violation occur in more general QFTs, particularly in Yang-Mills (YM) theories?

In this paper, we address this question and provide concrete evidence that pure YM theory is non-unitarity in $4-2\epsilon$ dimensions.
Due to the additional spin structure and gauge symmetry, YM theories are much more complicated than scalar theories, making this generalization highly non-trivial.
Our study is closely based on the recent construction of gluonic evanescent operators in \cite{Jin:2022ivc, Jin:2022qjc}.
We compute the Gram matrix of gluonic operators and find that negative-norm states are associated with the YM evanescent operators.
Moreover, these operators lead to complex anomalous dimensions, which we demonstrate in the one-loop order.
Our new results for the YM theory thus suggest that unitarity violation should be ubiquitous in QFT at non-integer spacetime dimensions.

Compared to the scalar theory case, an important new feature is that both negative-norm states and complex anomalous dimensions appear at dimension-12 operators in the YM theory (here we consider operators that are Lorentz scalars).
In contrast, in scalar theory, negative-norm states start at dimension-15 operators and complex anomalous dimensions appear at much higher dimension-23 \cite{Hogervorst:2015akt}.
Also, unlike scalar theory having an infrared (IR) fixed point, pure YM theory is asymptotically free and is expected to have an ultraviolet (UV) fixed point at spacetime dimension $4-2\epsilon$, see \emph{e.g.}~\cite{Peskin:1980ay, Gies:2003ic, Morris:2004mg}. Our computation shows that there are negative-norm evanescent states appearing for $4< d < 5$. Interestingly, the number of negative-norm states precisely matches the number of pairs of complex anomalous dimensions for the length-4 operators (operators containing four $F_{\mu\nu}$'s) up to mass dimension 16.

We would like to note that the use of dimensional regularization should not affect the unitarity of gauge theories in four dimensions.
Although the spacetime dimension is analytically continued to $d=4-2\epsilon$ during computations to regularize divergences, the limit as $\epsilon$ approaches zero is always taken, and one is interested in the finite physical observables. 
In this four-dimensional limit, only physical (namely non-evanescent) eigenstates are related to observables. Such states always have positive norms and real anomalous dimensions. 

Below, we first give a brief review of the gluonic evanescent operators. Then we discuss the negative-norm states and complex anomalous dimensions. We also explore a generalization to the theory of YM coupled to scalars, which can have an IR fixed point. Finally, we comment on the implications of the unitarity violation as well as some generalizations.

\section{Gluonic evanescent operators}
\label{sec:evanescentoper}

\noindent
In the pure YM theory, local gauge invariant operators are constructed by taking products of the field strength $F_{\mu\nu}$ and covariant derivatives $D_\mu$.
The simplest operator is the term ${\rm tr}(F_{\mu\nu}F^{\mu\nu})$ in the Lagrangian, which has a mass dimension of four.
A special class of operators is the so-called \emph{evanescent operators}, which are defined in general $d$ dimensions but vanish in the limit $d\rightarrow 4$.
Evanescent operators can be constructed by multiplying a tensor operator by a
Kronecker delta $\delta^{\mu_1\cdots\mu_n}_{\nu_1\cdots\nu_n}$
with $n\geq5$ and then taking Lorentz contractions.
For example,
\begin{equation}
\label{eq:deltaeg}
\mathcal{O}_{\rm a}=
\delta^{\mu_1\mu_2\mu_3\mu_4\mu_5}_{\nu_1\nu_2\nu_3\nu_4\nu_5}
\mathrm{tr}( D_{\nu_5} F_{\mu_1\mu_2}
D_{\mu_5} F_{\nu_1\nu_2}
F_{\mu_3\mu_4}F_{\nu_3\nu_4}).
\end{equation}
For convenience,
we will refer to the non-evanescent operators as physical operators.

In this paper, we focus on operators that are Lorentz scalars with all Lorentz indices contracted.
Because we study operators that are covariant in $d$-dimensional spacetime, the Lorentz contractions do not involve Levi-Civita $\epsilon$-tensor; thus, all operators are parity-even.

Gluonic evanescent operators begin to appear at a canonical dimension of 10 and involve at least four $F_{\mu\nu}$'s, such as the one in \eqref{eq:deltaeg}.
For the purposes of later computations, it is important to classify operators according to their mass dimension $\Delta_0$.
Two operators are said to be \emph{equivalent} if their difference is proportional to the equation of motion ($D_\mu F^{\mu\nu} = 0$) or  Bianchi identity ($D_\mu F_{\nu\rho}+D_\nu F_{\rho\mu}+D_\rho F_{\mu\nu} = 0$).
One can construct the basis of operators at a given mass dimension by eliminating such equivalence.
We keep operators with total derivatives in the operator basis.
A systematic classification of gluonic evanescent operators has been studied in \cite{Jin:2022ivc}.

\begin{table}[!t]
\centering
\begin{tabular}{|c|c|c|c|c|c| }
\hline
$\Delta_0$ & $~8~$ & $ ~10~$ & $ 12$ & $ 14$ & $ 16$\\
\hline
$N^{\rm p}$  & 4 & 20 & 82 & 232 & 550  \\
\hline
$N^{\rm e}$  & 0 & 4 & 25 & 92 & 259  \\
\hline
\end{tabular}
\caption{\label{tab:counting-operators} Counting of single-trace length-4 basis up to dimension $16$. $N^{\rm p}$ and $N^{\rm e}$ are the numbers of physical and evanescent operators, respectively.}
\end{table}

For the computation of the Gram matrix and the one-loop anomalous dimensions, we note that there is no mixing between operators containing different numbers of $F_{\mu\nu}$.
In this work, we will mainly focus on single-trace length-4 operators
(those containing four $F_{\mu\nu}$ but an arbitrary number of $D_\mu$),
which will be sufficient for the study of unitarity violation.
In Table~\ref{tab:counting-operators}, we summarize the count of length-4 basis operators (including both evanescent and physical operators) up to dimension $16$.

\section{Negative norms}
\label{sec:negativenorm}

\noindent
In this section, we demonstrate that the Hilbert space of local operators is not positive definite in the YM theory in non-integer dimensions.
This issue can be examined by considering the Gram matrix, which is a symmetric matrix defined as $G_{ij}$ in the two-point Green function:
\begin{align}
\label{eq:def}
\langle O_i(x)O_j(0)\rangle=\frac{G_{ij}}{|x^2|^{\Delta_{i}}}\,.
\end{align}
We will show that the Gram matrix is not positive definite, indicating the presence of negative-norm states, which implies that the theory is non-unitary.

For simplicity, we compute the Gram matrix in the free YM theory by setting the YM coupling $g=0$.
This provides the leading-order result of the Gram matrix.

For Lorentz-invariant operators,  Gram matrices depend only on the rank of gauge group $N_c$, the spacetime dimension $d$, and the canonical dimensions of the operators.
As a simple example, the norm of the dimension-4 operator ${\rm tr}(F_{\mu\nu}F^{\mu\nu})$ is
\begin{align}
G_{\mathrm{tr}F^2,\mathrm{tr}F^2}
=8\, N_c^2  \,d(d-1)(d-2)^3\,.
\end{align}

A notable feature of Gram matrices is the appearance of negative norms at $d=4-2\epsilon$ for $\epsilon<0$.
The lowest canonical dimension allowing for negative-norm states is $\Delta_0=12$.
Consider dimension-12 length-4 single-trace operators as an example.
This set contains 107 independent operators (see Table~\ref{tab:counting-operators}),
and the corresponding Gram matrix has a negative eigenvalue, thus not being positive definite.
To illustrate the emergence of negative-norm states, consider the following operator
\begin{equation}
\label{eq:opd12delta6}
\mathcal{O}_{\rm b}
=
\partial_{\rho} \partial_{\sigma}\big[\delta^{\mu_1\mu_2\mu_4\mu_5\nu_1\rho}_{
\mu_3\nu_2\nu_3\nu_4\nu_5\sigma}\text{tr}(D_{\mu_1}F_{\mu_2\mu_3}F_{\mu_4\mu_5}
D_{\nu_1}F_{\nu_2\nu_3}F_{\nu_4\nu_5})
\big].
\end{equation}
This operator contains a rank-6 Kronecker delta, so
it vanishes at $d=4$ and $d=5$.
Explicitly, the norm of $\mathcal{O}_{\rm b}$ is
\begin{align}
\label{eq:opd12delta6b}
G_{\rm bb}&= 1152\, N_c^2(N_c^2-1)   (3d+8)(d+2)(d+1)d^5
\nonumber\\
& \qquad \times (d-1)^2 (d-2)^5(d-3)(d-4)(d-5)\,,
\end{align}
which is negative when $4<d<5$.
Therefore, the Gram matrix for the $\Delta_0=12$ operators is non-positive definite at $d=4-2\epsilon$ ($\epsilon<0$).
As a comparison, for operators containing a rank-5 Kronecker delta,
their norms lack the $(d-5)$ factor and remain positive therein.

The operator example ${\cal O}_{\rm b}$ also suggests that
negative-norm states exist for operators with arbitrarily higher canonical dimensions.
Such operators can be constructed by
inserting arbitrary pairs of contracted $D_\mu$ into the dimension-12 operator (\ref{eq:opd12delta6}).
The norm of such an operator also has a factor linear in $(d-4)$ and $(d-5)$ and thus is negative when $4<d<5$.

We have computed Gram matrices for higher-dimension bases and found that the number of negative-norm states increases with $\Delta_0$.
In Table~\ref{tab:counting} we present the numbers of positive- and
negative-norm states for single-trace length-4 operators up to to dimension-16,
counted from the positive and negative eigenvalues of the Gram matrices.
The number of negative eigenvalues
always equals the number of linearly independent evanescent operators containing
rank-6 Kronecker deltas.

\begin{table}[!t]
\centering
\begin{tabular}{|c|c|c|c|c|c| }
\hline
$\Delta_0$ & $~8~$ & $ ~10~$ & $ 12$ & $ 14$ & $ 16$\\
\hline
\hline
$N_+^{\rm p}=N^{\rm p}$  & 4 & 20 & 82 & 232 & 550  \\
\hline
$N_+^{\rm e}$  & 0 & 4 & 24 & 88 & 246  \\
\hline
$N_-^{\rm e}$  & 0 & 0 & 1 & 4 & 13  \\
\hline
$N_{\gamma\textrm{-complex}}$  & 0 & 0 & $1\times2$ & $4\times2$ & $13\times2$  \\
\hline
\end{tabular}
\caption{\label{tab:counting} Counting of states with positive and negative norms for the single-trace length-4 basis up to mass dimension $16$.
$N_+$ and $N_-$ are the numbers of positive- and negative-norm states, respectively.
$N_{\gamma\textrm{-complex}}$ is the number of complext anomalous dimensions.}
\end{table}

To better understand the features of the YM evanescent operators,
let us compare them with those in scalar theory.
In YM theory, both the field strengths $F_{\mu\nu}$ and covariant derivatives $D_\mu$ carry Lorentz indices,
while in scalar theory, all Lorentz indices come solely from the covariant derivatives.
As a result, evanescent operators
in scalar theory must have length $L\geq5$ (that is, they must contain at least five scalar fields).
This explains why negative-norm states appear at higher-dimensional operators in scalar theory; for example,
in \cite{Hogervorst:2015akt}, it was found that the first Gram matrix block containing both negative- and positive-norm evanescent states appears at a canonical dimension $\Delta_0 =18$ and length $L=6$ in scalar theory.
In contrast, in YM theory, the Lorentz structure of the operators is much richer, and as discussed above, the similar Gram matrix with negative-norm states first appears at dimension 12 and with length $L=4$.

So far, we have considered the Gram matrix in the ``free" YM theory with $g=0$. This leading-order result  is expected to capture the main feature. In the next section, we will consider the interaction theory and compute the anomalous dimensions, and we will find the unitarity-violating effect is consistent with the above free theory results.

\section{Complex anomalous dimensions}
\label{sec:complexAD}

\noindent
The scaling dimension $\Delta$ of a local operator is defined as the sum of its canonical (classical) dimension $\Delta_0$ and a quantum correction part $\gamma$, known as the anomalous dimension (AD).
In a unitary CFT, the spectrum of scaling dimensions should always be real and bounded from below.
In this section, we will show that gluonic evanescent operators have complex ADs at the WF fixed point of the YM theory, thus manifestly demonstrating the violation of unitarity.

We first briefly review how to compute the ADs of local operators at the WF fixed point.
For a set of bare operators basis $\{O_i\}$ with the same canonical dimension $\Delta_0$, one can define the renormalized operators $\{O_{i,\text{R}}\}$ as
\begin{align}
	O_{i}^{\text{R}}=Z_i^{\ j}O_j\,,
\end{align}
where the renormalization matrix element $Z_i^{\ j}$ represents the mixing from $O_i$ to $O_j$ and is determined by UV poles in $\epsilon$ in the minimal subtraction scheme. The dilatation matrix is defined as
\begin{align}
	\label{dilatation}
	\mathcal{D}\equiv -\mu\frac{\text{d}Z}{\text{d}\mu}Z^{-1}\,.
\end{align}
The eigenvalues of the dilatation matrix give the ADs, which can be expanded as
\begin{equation}
 \gamma = \sum_{l=1}\left(\frac{\alpha_s}{4\pi}\right)^l \gamma^{(l)}\,,
\end{equation}
where $\alpha_s$ is the renormalized YM coupling constant.

At the WF fixed point \cite{Wilson:1971dc}, the beta function vanishes for the special value of the coupling $\alpha_*(\epsilon)$ (we work in $d=4-2\epsilon$ dimensions), which at the lowest order is given as
\begin{equation}
\alpha_*(\epsilon) = - { 4\pi \epsilon \over \beta_0} + {\cal O}(\epsilon^2) \,.\label{alphastar}
\end{equation}
In terms of $\alpha_*$, the dilatation matrix and the ADs are expanded in $\epsilon$, and the leading-order expansion is
\begin{equation}
\gamma_* = \epsilon \gamma^{(1)}_* + {\cal O}(\epsilon^2)\,, \qquad \gamma_*^{(1)} =  -\frac{\gamma^{(1)}}{\beta_0} \,.
\end{equation}
For example, for the operator ${\rm tr}(F_{\mu\nu}F^{\mu\nu})$,
\begin{equation}
\gamma_{*, {\rm tr}(F^2)}^{(1)} = 2 \,.
\end{equation}
We emphasize that since YM theory is asymptotically free with $\beta_0=\frac{11 N_c}{3} >0$, one gets a UV fixed point which is at $d>4$, \emph{i.e.}~$\epsilon<0$.

Let us provide a brief description of the strategy we use to compute ADs; one can refer to \cite{Jin:2022ivc, Jin:2022qjc} for a detailed discussion.
We first calculate the bare loop form factor,\footnote{
We comment that we can apply the on-shell unitarity-cut method \cite{Bern:1994cg,Bern:1994zx,Britto:2004nc}.
In our computation, both the tree-level blocks as well as the helicity-sum operation are in the fully $d$-dimensional Lorentz covariant form.
In this way, the obtained results are equivalent to the Feynman diagram computation which is applicable to non-unitary theories.
}
which is defined as
\begin{equation}
\mathcal{F}_{O}(1,..,n;q)=\int \text{d}^dx e^{-{\text{i}}q\cdot x}\langle \text{g}_1,\dots,\text{g}_n |O(x)|0\rangle\,.
\end{equation}
Next, we subtract the IR divergences according to Catani's formulas \cite{Catani:1998bh} to get the UV divergence. The UV divergence has a structure as a linear combination of the tree-level form factors of the basis operators, and the coefficients are matrix elements $Z_i^{\ j}$. Finally, we calculate the dilatation matrix according to the definition \eqref{dilatation}, whose eigenvalues give ADs.

We now present the results of the ADs.
Since there is no evanescent-to-physical operator mixing at one loop (namely, $(Z^{(1)})_{\rm e}^{~{\rm p}}=0$) \cite{Jin:2022ivc},
one can safely divide operators into evanescent and physical sectors and compute their one-loop ADs separately.
Our calculation shows that the one-loop complex ADs, which only occur in the evanescent sectors, begin to appear at canonical dimension 12.
This is consistent with the fact that negative-norm states start at this dimension, as discussed in the previous section.
The operator basis can be further classified into small sectors according to their parity under charge conjugation as well as Lorentz structures.
The $Z$ matrix will take a blockwise structure, and the ADs can be calculated within each sector.

The complex anomalous dimensions start to appear in a sector at dimension 12, where the operators are
\begin{align}
	&\partial_\nu \partial_\rho\Big[\delta^{12456\nu }_{3789\mu \rho }\Big(\text{tr}(D_{1}F_{23}F_{45}D_{6}F_{78}F_{9\mu })+\text{Rev.})\Big)\Big],\label{delta6D22}\\
	&\partial_\nu \partial_\rho\Big[\delta^{1}_{4} \delta^{2356\nu }_{789\mu \rho }\Big(\text{tr}(D_{1}F_{23}F_{45}D_{6}F_{78}F_{9\mu })+\text{Rev.})\Big)\Big],\nonumber\\
	&\partial_\nu \partial_\rho\Big[\delta^{1}_{4} \delta^{2356\nu }_{789\mu \rho }\Big(\text{tr}(D_{1}F_{23}D_{4}F_{56}F_{78}F_{9\mu })+\text{Rev.})\Big)\Big],\nonumber\\
	&\partial_\nu \partial_\rho\Big[\delta^{1}_{4} \delta^{2357\nu }_{689\mu \rho }\Big(\text{tr}(D_{1}F_{23}D_{4}F_{56}F_{78}F_{9\mu })+\text{Rev.})\Big)\Big],\nonumber\\
	&\partial_\nu \partial_\rho\Big[\delta^{1}_{4} \delta^{2367\nu }_{589\mu \rho }\Big(\text{tr}(D_{1}F_{23}F_{45}D_{6}F_{78}F_{9\mu })+\text{Rev.})\Big)\Big],\nonumber\\
	&\partial_\nu \partial_\rho\Big[\delta^{1}_{4} \delta^{2378\nu }_{569\mu \rho }\Big(\text{tr}(D_{1}F_{23}D_{4}F_{56}F_{78}F_{9\mu })+\text{Rev.})\Big)\Big],\nonumber\\
	&\partial_\nu \partial_\rho\Big[\delta^{1}_{5} \delta^{2347\nu }_{689\mu \rho }\Big(\text{tr}(D_{1}F_{23}D_{4}F_{56}F_{78}F_{9\mu })+\text{Rev.})\Big)\Big],\nonumber\\
	&\partial_\nu \partial_\rho\Big[\delta^{2}_{4}\delta^{1567\nu }_{389\mu \rho } \Big(\text{tr}(D_{1}F_{23}F_{45}D_{6}F_{78}F_{9\mu })+\text{Rev.})\Big)\Big]\,,\nonumber
\end{align}
where ``Rev." denotes reversion within the trace. Note that the first operator in \eqref{delta6D22} is the only dimension-12 operator containing a tensor degree-6 Kronecker symbol and is responsible for the existence of a negative-norm state.
We mention that we have focused on operators that are Lorentz scalars.
An alert reader may notice that the above operators are actually total derivatives of dimension-10, rank-2 tensor operators. We will see in \eqref{dim12eqs} below that the eigenvalue equation for these eight operators is not factorizable (with rational coefficients), which implies that their $Z$ matrix cannot be decomposed into smaller blocks. Thus, there should exist eight dimension-10, rank-2 primary operators that yield the same eight anomalous dimensions.

The one-loop dilatation matrix of this sector reads
\begin{align}
	N_c
	\left(
	\begin{array}{cccccccc}
		\frac{16}{3} & -2 & \frac{13}{12} & 0 & -\frac{14}{3} & 0 & -\frac{14}{3} & -\frac{28}{3} \\
		\frac{1}{2} & \frac{41}{6} & -2 & -\frac{5}{6} & \frac{2}{3} & \frac{5}{12} & \frac{7}{3} & \frac{16}{3} \\
		0 & 4 & 0 & -\frac{16}{3} & 0 & \frac{4}{3} & 0 & -\frac{16}{3} \\
		0 & \frac{4}{3} & -\frac{7}{3} & 4 & 0 & \frac{4}{3} & 0 & 0 \\
		-\frac{1}{12} & \frac{1}{12} & \frac{3}{8} & -\frac{1}{12} & \frac{22}{3} & -\frac{5}{8} & -\frac{1}{2} & -2 \\
		0 & -\frac{4}{3} & -\frac{2}{3} & 0 & 0 & \frac{32}{3} & 0 & \frac{16}{3} \\
		-\frac{1}{6} & -\frac{3}{2} & -\frac{9}{16} & \frac{1}{2} & -\frac{29}{6} & \frac{5}{12} & \frac{5}{6} & -\frac{13}{3} \\
		\frac{5}{6} & \frac{1}{3} & -\frac{13}{32} & \frac{5}{6} & -\frac{3}{4} & -\frac{1}{4} & -\frac{5}{12} & \frac{47}{6} \\
	\end{array}
	\right).
	\label{D22ad}
\end{align}
The one-loop ADs are given by the eigenvalue equation:
\begin{align}
	&x^8-\frac{257 x^7}{6}+\frac{27281 x^6}{36}-\frac{191654 x^5}{27}+\frac{3001838 x^4}{81}\nonumber\\
	&-\frac{24366124 x^3}{243}+\frac{21495296 x^2}{243}+\frac{101673536 x}{729}\nonumber\\
	&-\frac{175325696}{729}=0\,,
	\label{dim12eqs}
\end{align}
with $\gamma^{(1)}_*=-\frac{N_c x}{\beta_0}$.
Remarkably, two of the $\gamma^{(1)}_*$'s are complex, with numerical values:
\begin{align}
N_c(-1.8077 \pm 0.36228) \, \text{i} \,.
\end{align}
This provides further concrete evidence that pure YM theory is non-unitary in non-integer dimensions.\footnote{At one loop, complex ADs also exist for the theory in $4-2\epsilon$ dimensions away from the WF fixed point.}

We also compute the ADs for higher dimensional operators and find more complex ones.
For the length-4 operators up to $\Delta_0=16$, we observe an interesting pattern: the number of complex ADs is exactly twice the number of negative-norm states, which is summarized in Table~\ref{tab:counting}.
This match is not a general feature though; for example, it breaks down for the dimension-12 length-5 operators,
where there are 8 negative-norm states but only 7 pairs of complex ADs.
More details will be given in \cite{Jin:2023fbz}. 

Furthermore, the one-loop results already provide important implications for the properties at higher loops.
In particular, for a sector of operators that has no complex AD and also has no degeneracy of ADs at one loop, this sector will not have any complex AD at higher loop orders.
This can be understood by following a standard perturbative calculation in quantum mechanics, see \emph{e.g.} \cite{griffiths2018introduction}.

\section{Generalization with scalar fields}
\label{sec:scalarYM}

\noindent
As a generalization of pure YM theory, we consider YM theory coupled minimally to $N_f$ scalar fields:
\begin{align}
	{\cal L}_{\rm YMS} = {\cal L}_{\rm YM}+\sum_{i=1}^{N_f}\frac{1}{2}D^\mu\phi^a_i D_\mu\phi^a_i \,,
\end{align}
where $N_f$ is the number of flavors and the scalars are in the adjoint representation with $a=1,\dots,N_c^2-1$. 
The one-loop beta function differs from that of pure YM, and one should change the $\beta_0$ in \eqref{alphastar} to
\begin{align}
	\beta^{\text{YMS}}_0=\frac{11}{3}-\frac{N_f}{6}\,.
	\label{YMSbeta}
\end{align}
For $N_f=22$, the one-loop beta function vanishes, and for $N_f>22$, the fixed point becomes an IR one at $d<4$.

The operator sector \eqref{delta6D22} is enlarged by mixing with two new flavor-singlet operators\footnote{These operators are themselves C-even, thus there is no need for a ``Rev." term as in \eqref{delta6D22}.}
\begin{align}
	&\frac{1}{N_f}\sum_i\partial_\mu \partial_\nu\Big[\delta^{1245\nu}_{3678\mu }\Big(\text{tr}(D_{1}\phi_i D_{2, 3}\phi_i D_{4}F_{56}F_{78})\Big)\Big]\,,\label{YMSscalar}\\
	&\frac{1}{N_f}\sum_i\partial_\mu \partial_\nu\Big[\delta^{1235\nu}_{4678\mu }\Big(\text{tr}(D_{1}\phi_i D_{2}F_{34}D_{5, 6}\phi_i F_{78})\Big)\Big]\,.\nonumber
\end{align}
The one-loop dilatation matrix \eqref{D22ad} is expanded accordingly as
\begin{small}
\begin{align}
	N_c 
	\left(
	\begin{array}{c|c}
		\text{\Large{$\frac{{\rm Eq.}\eqref{D22ad}}{N_c}$}}
		& 
		\begin{array}{cc}
			-16 N_f & 0 \\
			8 N_f & 0   \\
			-\frac{32 N_f}{3} & 0 \\
			0 & 0 \\
			-4 N_f & 0 \\
			\frac{32 N_f}{3} & 0 \\
			-\frac{20 N_f}{3} & 0 \\
			-\frac{10 N_f}{3} & 0 
		\end{array}
		\\ \hline
		\begin{array}{cccccccc}
			0 & 0 & \frac{5}{192} & 0 & -\frac{5}{24} & 0 & -\frac{5}{24} & -\frac{5}{12}\\
			0 & 0 & 0 & 0 & 0 & 0 & 0 & 0
		\end{array}
		&
		\begin{array}{cc}
			\frac{N_f}{3}+8 & -\frac{4}{3} \\
			-\frac{2}{3} & \frac{N_f+22}{3}
		\end{array}
	\end{array}
	\right).
\end{align}
\end{small}
It is not difficult to show that, for any positive integer $N_f$, there is always one pair of complex ADs in this operator sector, see \cite{Jin:2023fbz} for details.

\section{Discussion}
\label{sec:discussion}

\noindent
In this study, we demonstrate that YM evanescent operators yield negative-norm states and give rise to complex anomalous dimensions.
Our findings indicate that the YM theory exhibits non-unitarity in fractional spacetime dimensions.
This observation, previously noted only for scalar theories \cite{Hogervorst:2014rta, Hogervorst:2015akt}, greatly extends our understanding of the theoretical landscape.
While the pure YM theory is expected to have a UV conformal fixed point at $d>4$, 
one can couple YM with a sufficiently large number of matter fields such that the theory has an IR conformal fixed point at $d<4$ (see \emph{e.g.}~\cite{Banks:1981nn}).
We have considered such a YM-scalar theory and show that the complex anomalous dimensions persist.
It would be interesting to check this for more general gauge theories.

Besides the evidence of negative-norm states and complex anomalous dimensions, it would be worthwhile to explore if the unitarity-violating effects could be manifested in other observables, such as correlation functions or the S-matrix.\footnote{We note that in special theories, it may be possible for negative norm states to lead to a unitary S-matrix, as in the Lee-Wick model \cite{Lee:1969fy}. However, the pure YM theory we consider is certainly different.}
The unitarity violation makes it questionable to apply the standard conformal bootstrap method \cite{Rattazzi:2008pe} to general CFTs in non-integer dimensions.
Note that the bootstrap method may provide good approximations in some cases~\cite{El-Showk:2013nia} (see also \cite{Codello:2014yfa, Golden:2014oqa, Chester:2014gqa}),
and this may be explained by the fact that unitarity violation effects occur at relatively high-dimensional states, such that the unitarity-violating effect is suppressed \cite{Hogervorst:2014rta, Hogervorst:2015akt}.
In the YM theory, because operators with negative norms or complex anomalous dimensions appear at lower dimensions than in the scalar theory, we expect that the unitarity-violating effects of the YM theory are stronger than those in the scalar theory.
We mention that other bootstrap methods that do not rely on unitarity (see \emph{e.g.}~\cite{Gliozzi:2013ysa}) should still work in this case.

It would be interesting to generalize our study to gravitational theories, in which similar evanescent operators can be defined, for example, by replacing $F_{\mu\nu}$ with $R_{\mu\nu}$.
Note that local operators are not physical observables in gravity (as they break diffeomorphism invariance), and one may need to consider other observables such as amplitudes (see \emph{e.g.}~\cite{Bern:2015xsa, Bern:2017puu}) or non-local operators.

Our findings have significant implications when viewed through the lens of holographic duality \cite{tHooft:1993dmi, Susskind:1994vu, Maldacena:1997re}, where a gauge theory in $d$ dimensions is holographically related to a gravity or string theory in $d+1$ dimensions. 
The non-unitary gauge theory in fractional $d$ dimensions thus implies that its gravitational counterpart also violates unitarity. 
Unraveling how these violations manifest in the gravitational realm could prove to be a fascinating and insightful endeavor.

\vskip .3cm
{\it Acknowledgments.}
We would like to thank Bo Feng, Yunfeng Jiang, Jianxin Lu, Tao Shi, and Jun-Bao Wu for the discussion.
This work is supported in part by the National Natural Science Foundation of China (Grants No.~12175291, 11935013, 12047503) and by the CAS under Grants No.~YSBR-101.
We also thank the support of the HPC Cluster of ITP-CAS.


\end{document}